\documentclass[a4paper,11pt]{article}
\usepackage[dvips]{graphicx}
\usepackage{amssymb}
\usepackage{amsmath}

\usepackage{cite}

\begin{document}

\title{Supersymmetric Chern-Simons Vortex Systems and Extended Supersymmetric Quantum Mechanics Algebras}
\author{V. K. Oikonomou\thanks{
voiko@physics.auth.gr}\\
Max Planck Institute for Mathematics in the Sciences\\
Inselstrasse 22, 04103 Leipzig, Germany} \maketitle

\begin{abstract}
We study $N=2$ supersymmetric Chern-Simons Higgs models in $(2+1)$-dimensions and the existence of extended underlying supersymmetric quantum mechanics algebras. Our findings indicate that the fermionic zero modes quantum system in conjunction with the system of zero modes corresponding to bosonic fluctuations, are related to an $N=4$ extended 1-dimensional supersymmetric algebra with central charge, a result closely connected to the $N=2$ spacetime supersymmetry of the total system. We also add soft supersymmetric terms to the fermionic sector in order to examine how this affects the index of the corresponding Dirac operator, with the latter characterizing the degeneracy of the solitonic solutions. In addition, we analyze the impact of the underlying supersymmetric quantum algebras to the zero mode bosonic fluctuations. This is relevant to the quantum theory of self-dual vortices and particularly for the symmetries of the metric of the 
space of vortices solutions and also for the non-zero mode states of bosonic fluctuations.
\end{abstract}

\section*{Introduction}

Gauge field theories in $(2+1)$-dimensions \cite{dunnebook,lee} have particularly interesting attributes, mainly stemming from the fact that the Chern-Simons \cite{lee1,lee1a,lee1b} term can be consistently incorporated in the theoretical framework. This term significantly alters the long distance behavior of the theory and also modifies the characteristics of the allowed solitonic solutions. Particularly, the topologically stable solitonic solutions can have electric charge \cite{lee3,lee3a,lee3b}, a feature that is absent in the usual $(2+1)$-dimensional Abelian Higgs model. For specific choices of the Higgs potential, the minimum energy static vortex solutions satisfy a set of first-order differential self-duality equations, the Bogomol'nyi equations \cite{dunnebook,lee19,lee7,lee16}. These self-dual systems have static multisoliton solutions that can be localized in different points in space, or superimposed at one point.

\noindent Supersymmetric extensions of $(2+1)$-dimensional gauged Chern-Simons models have interestingly appealing features \cite{lee}, with the most interesting of these features, related to the fermionic and bosonic zero modes, in the background of these self-dual vortices \cite{lee}. The fermionic zero modes are relevant to the quantum theory of these models, since these represent the degeneracy of the solitonic states, while the zero modes of the bosonic fluctuations determine the collective coordinates that describe the solitons and their small velocity kinematics. In some models, especially in $N=2$ spacetime supersymmetric models, the fermionic zero modes are directly related to the zero modes of bosonic fluctuations, with the latter describing massless modes around the vortices (see for example \cite{lee}, and references therein).

\noindent Since the aforementioned zero modes play a crucial role towards the formulation of the quantum theory of supersymmetric Chern-Simons vortices, in this paper we shall present an interesting property of the zero modes of models with global $N=2$ spacetime supersymmetry. Particularly, we find that the zero modes of fermionic and bosonic fluctuations of Abelian gauge models having a global $N=2$ spacetime supersymmetry in $(2+1)$-dimensions, can separately constitute two $N=2$, $d=1$ supersymmetric quantum algebras \cite{witten,susyqm,susyqm1,susyqm2,susyqm3,susyqm4,susyqm5,thaller}, with the zero modes being the corresponding quantum Hilbert space vectors. Moreover, due to the $N=2$ global spacetime supersymmetry of the theory, the underlying one dimensional $N=2$ supersymmetric quantum algebras can form a centrally extended $N=4$, $d=1$ supersymmetric quantum algebra. Hence, the quantum system that consists of fermionic zero modes and of zero modes corresponding to the
bosonic fluctuations, constitute an quantum system with centrally extended supersymmetry. For the bosonic fluctuations subsystem, this quantum algebra has important implications for the quantum theory of the vortices. This is owing to the fact that, the bosonic fluctuations correspond to collective coordinates describing the vortices positions and slow velocity kinematics. As we shall see, the underlying $N=2$, $d=1$ supersymmetric quantum mechanics algebra divides the Hilbert space of the excited states corresponding to the bosonic fluctuations, in two graded Hilbert spaces.

\noindent We shall study the models in two important limiting cases of the theory, namely for $\kappa =0$ and $\kappa \rightarrow 0$, with $\kappa$ the coupling of the Chern-Simons term. The first case, that is the $\kappa =0$ case, corresponds to the supersymmetric extension of the Landau-Ginzburg model. Hence, our results could have interesting applications to quantum Hall effect and to high-temperature superconductivity \cite{oh,oh1,oh2}. Inspired from these applications, we introduce some soft supersymmetry breaking terms to the fermionic sector of the global $N=2$ spacetime supersymmetric quantum model, to see to what extend the zero modes of the fermionic sector are affected. We shall see that the index of the corresponding Dirac operator remains intact. These soft supersymmetry breaking terms are similar to terms introduced by inhomogeneities in superconductors, so this result has an intrinsic appeal, since the fermionic zero modes represent the degeneracy of the solitonic states. Thereby, as we shall 
see,
the degeneracy of the solitonic solutions for a $(2+1)$-dimensional supersymmetric Landau-Ginzburg model is robust to inhomogeneities quantified in terms of soft supersymmetry breaking terms.

\noindent As we already mentioned, the bosonic fluctuations determine the metric of the space that the vortices are located and are free to move. This metric has a global $SO(2)$ rotational symmetry, a feature that is also met in the underlying $N=2$, $d=1$ quantum structure. Actually, we shall demonstrate that the supercharges of the underlying supersymmetric quantum mechanics algebra have a global $U(1)$ supersymmetry which is transformed to an  $SO(2)$ when the supercharges are real. This global rotational symmetry is inherited to the Hilbert space states of the aforementioned quantum system, which are the zero modes of the bosonic fluctuations, thus indirectly establishing that the rotational symmetry of the metric is not accidental, but has its origin to the underlying $N=2$, $d=1$ quantum algebra (shortened to SUSY QM hereafter).

\noindent In contrast to the global $N=2$ spacetime supersymmetric model, the $N=1$ model does not lead to an extended supersymmetric algebra between the fermionic and bosonic fluctuations zero modes, but only leads to two different $N=2$, $d=1$ supersymmetries, one for each subsystem.

\noindent This paper is organized as follows. In order to render the paper self contained, in section 1, we briefly review the supersymmetric Chern-Simons Abelian models in $(2+1)$-dimensions. In section 2, we analyze in full detail the supersymmetric quantum mechanics algebras that underlie the bosonic and fermionic sector of the models for the $\kappa =0$ case. Moreover, we present the $SO(2)$ invariance of the metric, with the latter describing the space that the vortices are located, and also we explicitly show that the same symmetry governs the $N=2$, $d=1$ algebra that underlies the bosonic fluctuations. The soft supersymmetric fermionic terms and their effect on the index of the Dirac operator, are studied in section 3. In section 4, we study the $\kappa \rightarrow \infty$ case of the Chern-Simons supersymmetric model and additionally, the corresponding model for the case of $N=1$ global spacetime supersymmetry. Finally, the conclusions follow in the end of the paper.

\section{Supersymmetric Maxwell-Chern-Simons Theory}
Global supersymmetry, and particularly $N=2$ spacetime supersymmetry plays a very crucial role in the existence of static vortex solutions. This is due to the fact that in the non-supersymmetric $(2+1)$-dimensional Abelian Higgs model with both Maxwell and Chern-Simons terms in the action, self dual static vortex solutions occur only for specific forms of the scalar potential. If spacetime supersymmetry is imposed, then the potential form is uniquely determined \cite{lee}. Additionally to the $N=2$ supersymmetry, the $N=1$ spacetime supersymmetric extension, exactly reproduces the bosonic spectrum of the $N=2$ theory. The latter situation is particularly interesting since within the $N=2$ framework, the fermionic zero modes are directly related to the bosonic zero modes, something that is absent in the $N=1$ spacetime supersymmetric extension. This fact has particularly interesting consequences, with respect to the underlying one dimensional supersymmetries. We now present the models that we will extensively 
use in the following, based on reference \cite{lee}.
\noindent The Lagrangian for the $N=2$ spacetime supersymmetric model is \cite{lee},
\begin{align}\label{n2lag}
&\mathcal{L}=-\frac{1}{4}F_{\mu \nu}^{\mu \nu}+\frac{1}{4}\kappa \epsilon^{\mu \nu \lambda}F_{\mu \nu}A_{\lambda}-\lvert D_{\mu}\phi \lvert^{2}-\frac{1}{2}(\partial_{\mu}N)^2
\\ \notag & -\frac{1}{2}(e\lvert\phi\lvert^2+\kappa N-eu^2)^2-e^2N^2\lvert\phi\lvert^2
\\ \notag & i\bar{\psi}\gamma^{\mu}D_{\mu}\psi+i\bar{\chi}\gamma^{\mu}\partial_{\mu}\chi+\kappa\bar{\chi}\chi
+i\sqrt{2}e(\bar{\psi}\chi\phi-\bar{\chi}\psi\phi^*)+e N\bar{\psi}\psi
\end{align}
where $D_{\mu}=\partial_{\mu}-ieA_{\mu}$ and the gamma matrices satisfy the relation $\gamma^{\mu}\gamma^{\nu}=-\eta^{\mu\nu}-i\epsilon^{\mu \nu \lambda}\gamma_{\lambda}$. The fields $\psi ,\chi$ are two component spinors, with $\psi$ being a Weyl charged fermionic field and $\chi$ a neutral complex Weyl two component spinor.
The corresponding supersymmetry transformations, under which the Lagrangian (\ref{n2lag}) is invariant, are,
\begin{align}\label{susytrans}
& \delta_{\eta}A_{\mu}=i(\bar{\eta}\gamma_{\mu}\chi-\bar{\chi}\gamma_{\mu}\eta),
\\ \notag & \delta_{\eta}\phi=\sqrt{2}\bar{\eta}\psi,{\,}{\,}{\,}\delta_{\eta}N=i(\bar{\chi}\eta-\bar{\eta}\chi),
\\ \notag & \delta_{\eta}\psi=-\sqrt{2}(i\gamma^{\mu}\eta D_{\mu}\phi-\eta e N\phi),
\\ \notag & \delta_{\eta}\chi=\gamma^{\mu}\eta(\partial_{\mu} N+\frac{1}{2} i\epsilon_{\mu \nu \lambda}F^{\nu \lambda})+i\eta(e\lvert \phi \lvert^2+\kappa N-e v^2).
\end{align}
with $\eta$ an auxiliary complex Grassmannian variable.

\noindent For later convenience, we present here the $N=1$ spacetime supersymmetric Lagrangian. This is equal to:
\begin{align}\label{n2lagf}
&\mathcal{L}'=-\frac{1}{4}F_{\mu \nu}^{\mu \nu}+\frac{1}{4}\kappa \epsilon^{\mu \nu \lambda}F_{\mu \nu}A_{\lambda}-\lvert D_{\mu}\phi \lvert^{2}-\frac{1}{2}(\partial_{\mu}N)^2
\\ \notag & -\frac{1}{2}(e\lvert \phi \lvert^2+\kappa N-eu^2)^2-e^2 N^2\lvert \phi\lvert^2
\\ \notag & i\bar{\psi}\gamma^{\mu}D_{\mu}\psi+i\bar{\chi}\gamma^{\mu}\partial_{\mu}\chi+\frac{1}{2}\kappa (\bar{\chi}\chi^{c}+\bar{\chi}^{c}\chi)-i\sqrt{2}e(\bar{\psi}\chi^c\phi-\bar{\chi}^c\psi\phi^*)-e N\bar{\psi}\psi
\end{align}
Accordingly, the $N=1$ supersymmetry transformations are given by:
\begin{align}\label{susytrans1}
& \delta_{\eta}\psi=-\sqrt{2}(i\gamma^{\mu}\eta D_{\mu}\phi+\eta e N\phi),
\\ \notag & \delta_{\eta}\chi=\gamma^{\mu}\eta(\partial_{\mu} N+\frac{1}{2} i\epsilon_{\mu \nu \lambda}F^{\nu \lambda})-i\eta(e\lvert \phi \lvert^2+\kappa N-e v^2).
\end{align}
with $\eta$ a Majorana spinor. Note that the bosonic part of the $N=1$ Lagrangian coincides with the bosonic part of the $N=2$ Lagrangian.

\subsection{Fermionic Zero Modes of the $N=2$ Supersymmetric System}

We now turn our focus on the fermionic part of the $N=2$ Lagrangian in order to find the fermionic zero modes, in the background of self dual vortices. The fermionic equations of motion corresponding to Lagrangian (\ref{n2lag}), are:
\begin{align}\label{eqnmotion}
& \gamma^i D_i \psi+ie(\gamma^0 A^0-N)\psi-\sqrt{2}e\phi \chi=0
\\ \notag & \gamma^i\partial_i \chi -i\kappa \chi+\sqrt{2}e\phi^*\psi=0
\end{align}
By making the following conventions:
\begin{equation}\label{conventions}
\gamma^0=\sigma_3,{\,}{\,}\gamma^1=i\sigma_2,{\,}{\,}\gamma^2=i\sigma_1
\end{equation}
and by setting:
\begin{equation}\label{setting}
\psi=\left (\begin{array}{c}
        \psi_{\uparrow}          \\
    \psi_{\downarrow} \\
\end{array}\right ),{\,}{\,}{\,}\chi=\left (\begin{array}{c}
        \chi_{\uparrow}          \\
    \chi_{\downarrow} \\
\end{array}\right )
\end{equation}
and additionally assuming positive values of the flux, we obtain the equations of motion for the fermions:
\begin{align}\label{fermionsequationsofmotions}
&(D_1+iD_2)\psi_{\downarrow}-\sqrt{2}e\phi \chi_{\uparrow}=0
\\ \notag & (\partial_1-i\partial_2)\chi_{\uparrow }+i\kappa \chi_{\downarrow }-\sqrt{2}\phi^*\psi_{\downarrow }=0
\\ \notag & (D_1-iD_2)\psi_{\uparrow }+2i e A^0\psi_{\downarrow }+\sqrt{2}e\phi \chi_{\downarrow }=0
\\ \notag & (\partial_1+i\partial_2)\chi_{\downarrow }-i\kappa \chi_{\uparrow }+\sqrt{2}e\phi^*\psi_{\uparrow }=0
\end{align}
We shall investigate the theory for the two limiting cases of the Chern-Simons parameter $\kappa$, namely, for $\kappa =0$ and $\kappa \rightarrow \infty$. Note that, when the Chern Simons coupling $\kappa$ becomes zero, the $N=1$ and $N=2$ Lagrangians reduce to the $N=2$ Abelian Higgs model.

\subsection{Fluctuating Bosonic Zero Modes of the $N=2$ Supersymmetric System}

In order to extract information for the self dual vortices solutions and correspondingly for the bosonic zero modes, we focus our interest on the bosonic part of the Lagrangian (\ref{n2lag}). The theory possesses two ground states, namely the non-symmetric one, with $\lvert \phi \lvert =v$, $N=0$ and a symmetric one with $\phi =0,{\,}{\,}N=\frac{ev^2}{\kappa}$. Solutions of the topological soliton type exist in non-symmetric phase which has the following asymptotic behavior \cite{lee}:
\begin{equation}\label{boun1}
\lim_{r\rightarrow \infty} N(r)\rightarrow 0,{\,}{\,}{\,}\lim_{r\rightarrow \infty}\lvert \phi (r)\lvert \rightarrow v
\end{equation}
and additionally a quantized flux $\Phi=\pm \frac{2\pi n}{e}$.
In the symmetric Higgs phase, $\phi =0$, $N=ev^2/\kappa$, non-topological solutions exist with the following asymptotic behavior:
\begin{equation}\label{boun2}
\lim_{r\rightarrow \infty}N(r)\rightarrow \frac{ev^2}{\kappa}+\frac{\mathrm{const.}}{r^{2a}},{\,}{\,}{\,}\lim_{r\rightarrow \infty}\lvert \phi (r)\lvert \rightarrow \frac{\mathrm{const.}}{r^a}
\end{equation}
A compelling constraint to all static solutions is that they have to satisfy the Gauss law:
\begin{equation}\label{gausslaw}
\partial_{i}F^{i0}+\kappa F_{12}-ie(\phi^*D^0\phi-D^0\phi^*\phi)=0
\end{equation}
Integrating over the whole space, we obtain a static configuration of magnetic flux $\Phi=\int \mathrm{d}^2xF_{12}$, which has a total electric charge, $Q=-\frac{\kappa \Phi}{e}$. The energy of the configuration is bounded from below by the relation $E\geq ev^2\lvert \Phi\lvert $, and is saturated if the configurations satisfy the self duality equations:
\begin{align}\label{selfdualeqns}
&(D_1\pm iD_2)\phi=0
\\ \notag & F_{12}\pm(e\lvert \phi\lvert^2 +\kappa N-ev^2)=0
\\ \notag & A^0\mp N=0
\\ \notag & \partial_{i}F^{i0}+\kappa F_{12}-ie(\phi^*D^0\phi-D^0\phi^*\phi)=0
\end{align}

\noindent The quantum aspects are revealed when fluctuations of the fields around the vortex solutions are considered. The most interesting of those fluctuations are the zero mode fluctuations, which enjoy an elevated role among the various field fluctuations, owing to the fact that these zero modes correspond to vortex related collective coordinates (see \cite{lee} and references therein). Actually, since it is rather difficult to determine the positions of the vortices in the configuration space, by studying the zero modes fluctuations, we can have information of equal importance. The equations of the zero modes fluctuations are obtained by varying the self duality equations around the static classical vortex configuration, and can be cast in the following form:
\begin{align}\label{selfdualfluctuat}
&(D_1+iD_2)\delta \phi-ie\phi (\delta A_1+i \delta A_2)=0
\\ \notag &\partial_1\delta A_2-\partial_2 \delta A_1+e(\phi^*\delta \phi +\phi\delta \phi^*)+k\delta A^0=0
\end{align}
In addition, by varying the Gauss law constraint we obtain:
\begin{equation}\label{gausslawconstr}
(-\nabla^2+\kappa^2+2e^2\lvert \phi \lvert^2)\delta A^0+e(\kappa+2eA^0)(\phi^*\delta \phi +\phi\delta \phi^*)=0
\end{equation}
We set $\kappa =0$ in order to describe the Landau-Ginzburg vortex situation, as we did in the fermion case. In that case, one can consistently set $A^0=N=0$, just as we did in the fermionic case.

\section{The $\kappa =0$ Case}

\subsection{$N=2$, $d=1$ SUSY QM Algebra in the Fermionic Sector}

The fermionic equations of motion (\ref{fermionsequationsofmotions}) for $\kappa =0$ become,
\begin{align}\label{fer1}
&(D_1+iD_2)\psi_{\downarrow}-\sqrt{2}e\phi\chi_{\uparrow}=0
\\ \notag & (\partial_1-i\partial_2)\chi_{\uparrow}-\sqrt{2}e\phi^*\psi_{\downarrow}=0
\\ \notag & (D_1-iD_2)\psi_{\uparrow}+\sqrt{2}e\phi\chi_{\downarrow}=0
\\ \notag & (\partial_1+i\partial_2)\chi_{\downarrow}+\sqrt{2}e\phi^*\psi_{\uparrow}=0
\end{align}
The last two equations of relation (\ref{fer1}) have no solutions describing localized fermions, but only have some trivial solutions \cite{lee}. Conversely, the first two equations of relation (\ref{fer1}), have $2n$ solutions, with $n$ the vorticity number \cite{lee}. Hence, we can form an operator $\mathcal{D}_{LG}$, corresponding to the first two equations of (\ref{fer1}),
\begin{equation}\label{susyqmrn567m}
\mathcal{D}_{LG}=\left(%
\begin{array}{cc}
 D_1+iD_2 & -\sqrt{2}e\phi
 \\ -\sqrt{2}\phi^* & \partial_1-i\partial_2\\
\end{array}%
\right)
\end{equation}
which acts on the vector:
\begin{equation}\label{ait34e1}
|\Psi_{LG}\rangle =\left(%
\begin{array}{c}
  \psi_{\downarrow} \\
  \chi_{\uparrow} \\
\end{array}%
\right).
\end{equation}
Thereby, the first two equations of (\ref{fer1}) can be written as:
\begin{equation}\label{transf}
\mathcal{D}_{LG}|\Psi_{LG}\rangle=0
\end{equation}
The solutions of the above equation yield the zero modes of the operator $\mathcal{D}_{LG}$, but since as we already mentioned, the first two equations of (\ref{fer1}) have $2n$ normalized solutions, we can easily conclude that:
\begin{equation}\label{dimeker}
\mathrm{dim}{\,}\mathrm{ker}\mathcal{D}_{LG}=2n
\end{equation}
Moreover, the adjoint of the operator $\mathcal{D}_{LG}$, namely $\mathcal{D}_{LG}^{\dag}$, is equal to:
\begin{equation}\label{eqndag}
\mathcal{D}_{LG}^{\dag}=\left(%
\begin{array}{cc}
 D_1-iD_2 & \sqrt{2}e\phi
 \\ \sqrt{2}\phi^* & \partial_1+i\partial_2\\
\end{array}%
\right)
\end{equation}
and acts on the vector:
\begin{equation}\label{ait3hgjhgj4e1}
|\Psi_{LG}'\rangle =\left(%
\begin{array}{c}
  \psi_{\uparrow} \\
  \chi_{\downarrow} \\
\end{array}%
\right).
\end{equation}
The zero modes of the corresponding adjoint operator $\mathcal{D}_{LG}^{\dag}$, correspond to the solutions of the last two equations of (\ref{fer1}), with the obvious replacement $e\rightarrow-e$. Obviously, since the corresponding pair of equations has no normalized solutions, the corresponding kernel of the adjoint operator is null, that is:
\begin{equation}\label{dimeke1r11}
\mathrm{dim}{\,}\mathrm{ker}\mathcal{D}_{LG}^{\dag}=0
\end{equation}
The normalization condition for the solutions of (\ref{fer1}) is crucial for our analysis, since only for such solutions, the operator $\mathcal{D}_{LG}$ is Fredholm, a result that can be verified by (\ref{dimeker}) and (\ref{dimeke1r11}). The Fredholm property is an exceptional attribute of the system under study as we will see latter. The fermionic system in the self dual vortices background, possesses an unbroken $N=2$, $d=1$ supersymmetry. Indeed, the supercharges of this $N=2$, $d=1$ SUSY algebra are related to the operator $\mathcal{D}_{LG}$ and are equal to:
\begin{equation}\label{s7}
\mathcal{Q}_{LG}=\bigg{(}\begin{array}{ccc}
  0 & \mathcal{D}_{LG} \\
  0 & 0  \\
\end{array}\bigg{)},{\,}{\,}{\,}\mathcal{Q}^{\dag}_{LG}=\bigg{(}\begin{array}{ccc}
  0 & 0 \\
  \mathcal{D}_{LG}^{\dag} & 0  \\
\end{array}\bigg{)}
\end{equation}
while the quantum Hamiltonian of the quantum system is:
\begin{equation}\label{s11}
\mathcal{H}_{LG}=\bigg{(}\begin{array}{ccc}
 \mathcal{D}_{LG}\mathcal{D}_{LG}^{\dag} & 0 \\
  0 & \mathcal{D}_{LG}^{\dag}\mathcal{D}_{LG}  \\
\end{array}\bigg{)}
\end{equation}
These three elements of the algebra, satisfy the $d=1$ SUSY algebra:
\begin{equation}\label{relationsforsusy}
\{\mathcal{Q}_{LG},\mathcal{Q}^{\dag}_{LG}\}=\mathcal{H}_{LG}{\,}{\,},\mathcal{Q}_{LG}^2=0,{\,}{\,}{\mathcal{Q}_{LG}^{\dag}}^2=0
\end{equation}
The Hilbert space of the supersymmetric quantum mechanical system, $\mathcal{H}_{LG}$ is $Z_2$ graded by the operator $\mathcal{W}$, the Witten parity, an involution operator. This operator commutes with the total Hamiltonian,
\begin{equation}\label{s45}
[\mathcal{W},\mathcal{H}_{LG}]=0
\end{equation}
and  anti-commutes with the supercharges,
\begin{equation}\label{s5}
\{\mathcal{W},\mathcal{Q}_{LG}\}=\{\mathcal{W},\mathcal{Q}_{LG}^{\dag}\}=0
\end{equation}
Moreover, $\mathcal{W}$ satisfies the following identity,
\begin{equation}\label{s6}
\mathcal{W}^{2}=1
\end{equation}
a property inherent to projective operators. Let us see what kind of projective operator this involution is. As we already mentioned, the Witten parity $\mathcal{W}$, spans the total Hilbert space into subspaces which are $Z_2$ equivalent. Hence the total Hilbert space of the quantum system can be written as:
\begin{equation}\label{fgjhil}
\mathcal{H}=\mathcal{H}^+\oplus \mathcal{H}^-
\end{equation}
with the vectors that belong to the two subspaces $\mathcal{H}^{\pm}$, classified according to their Witten parity, to even and odd parity states, that is:
\begin{equation}\label{shoes}
\mathcal{H}^{\pm}=\mathcal{P}^{\pm}\mathcal{H}=\{|\psi\rangle :
\mathcal{W}|\psi\rangle=\pm |\psi\rangle \}
\end{equation}
In addition, the corresponding Hamiltonians of the $Z_2$ graded spaces are:
\begin{equation}\label{h1}
{\mathcal{H}}_{+}=\mathcal{D}_{LG}{\,}\mathcal{D}_{LG}^{\dag},{\,}{\,}{\,}{\,}{\,}{\,}{\,}{\mathcal{H}}_{-}=\mathcal{D}_{LG}^{\dag}{\,}\mathcal{D}_{LG}
\end{equation}
The operator $\mathcal{W}$, in the case at hand, can be represented in the following matrix form:
\begin{equation}\label{wittndrf}
\mathcal{W}=\bigg{(}\begin{array}{ccc}
  1 & 0 \\
  0 & -1  \\
\end{array}\bigg{)}
\end{equation}
In equation (\ref{shoes}) we introduced the operator $\mathcal{P}$, which makes the classification of the Hilbert vectors to even an odd states more transparent, since the eigenstates of $\mathcal{P}^{\pm}$ the eigenstates of which, $|\psi^{\pm}\rangle$, satisfy
the following relation:
\begin{equation}\label{fd1}
P^{\pm}|\psi^{\pm}\rangle =\pm |\psi^{\pm}\rangle
\end{equation}
Thereby, we call them positive and negative parity eigenstates, with ``parity'' referring to the $P^{\pm}$ operator, which is nothing else but the Witten parity operator. Deploying the representation (\ref{wittndrf}) for the Witten parity operator,
 the parity eigenstates can represented by the vectors,
\begin{equation}\label{phi5}
|\psi^{+}\rangle =\left(%
\begin{array}{c}
  |\phi^{+}\rangle \\
  0 \\
\end{array}%
\right),{\,}{\,}{\,}
|\psi^{-}\rangle =\left(%
\begin{array}{c}
  0 \\
  |\phi^{-}\rangle \\
\end{array}%
\right)
\end{equation}
with $|\phi^{\pm}\rangle$ $\epsilon$ $\mathcal{H}^{\pm}$. Turning back to the fermionic system at hand, it is easy to write the fermionic states of the system in terms of the SUSY quantum algebra. We already wrote down the supercharges, namely relation (\ref{s7}), hence it easy to identify that:
\begin{equation}\label{fdgdfgh}
|\Psi_{LG}\rangle =|\phi^{-}\rangle=\left(%
\begin{array}{c}
  \psi_{\downarrow} \\
  \chi_{\uparrow} \\
\end{array}%
\right),{\,}{\,}{\,}|\Psi_{LG}'\rangle =|\phi^{+}\rangle=\left(%
\begin{array}{c}
  \psi_{\uparrow} \\
  \chi_{\downarrow} \\
\end{array}%
\right)
\end{equation}
Therefore, the corresponding even and odd parity SUSY quantum states are:
\begin{equation}\label{phi5}
|\psi^{+}\rangle =\left(%
\begin{array}{c}
  |\Psi_{LG}'\rangle \\
  0 \\
\end{array}%
\right),{\,}{\,}{\,}
|\psi^{-}\rangle =\left(%
\begin{array}{c}
  0 \\
  |\Psi_{LG}\rangle \\
\end{array}%
\right)
\end{equation}
upon which, the Hamiltonian and the supercharges act. Supersymmetry is unbroken if the Witten index is a non-zero integer. The Witten index for Fredholm operators is defined to be:
\begin{equation}\label{phil}
\Delta =n_{-}-n_{+}
\end{equation}
with $n_{\pm}$ the number of zero
modes of ${\mathcal{H}}_{\pm}$ in the subspace $\mathcal{H}^{\pm}$, with the constraint that these are finitely many.

\noindent The case for which the Witten index is zero and also
if $n_{+}=n_{-}=0$, then supersymmetry is broken. Conversely, if $n_{+}=
n_{-}\neq 0$, the system has still an unbroken supersymmetry.

\noindent The Witten index is directly connected to the Fredholm index of the operator $\mathcal{D}_{LG}$, as can be seen in the following equations:
\begin{align}\label{ker1}
&\Delta=\mathrm{dim}{\,}\mathrm{ker}
{\mathcal{H}}_{-}-\mathrm{dim}{\,}\mathrm{ker} {\mathcal{H}}_{+}=
\mathrm{dim}{\,}\mathrm{ker}\mathcal{D}_{LG}^{\dag}\mathcal{D}_{LG}-\mathrm{dim}{\,}\mathrm{ker}\mathcal{D}_{LG}\mathcal{D}_{LG}^{\dag}=
\\ \notag & \mathrm{ind} \mathcal{D}_{LG} = \mathrm{dim}{\,}\mathrm{ker}
\mathcal{D}_{LG}-\mathrm{dim}{\,}\mathrm{ker} \mathcal{D}_{LG}^{\dag}
\end{align}
Recalling the results of equations (\ref{dimeker}) and (\ref{dimeke1r11}), the Witten index is equal to:
\begin{equation}\label{fnwitten}
\Delta =-2n
\end{equation}
Hence, the fermionic system in the self-dual Landau-Ginzburg vortices background with $N=2$ spacetime supersymmetry, has an unbroken $N=2$, $d=1$ supersymmetry. This result could stem from the fact that the initial system has an unbroken $N=2$ spacetime supersymmetry, so the Hilbert space of the zero modes states retains a $N=2$, $d=1$ supersymmetric quantum algebra. Oddly this is not true, as we shall discuss in the following sections.

\noindent We must be cautious when we study supersymmetric algebras in different dimensions. The spacetime supersymmetric algebra is four dimensional while the SUSY quantum mechanics algebra is one dimensional. Moreover, spacetime supersymmetry in $d>1$ dimensions and
supersymmetric quantum mechanics, that is $d=1$ supersymmetry, are not the same. However, there is an obvious connection, since extended (with $N = 4, 6...$) supersymmetric quantum mechanics models can describe
the dimensional reduction to one (temporal) dimension of N = 2 and N = 1
Super-Yang Mills models \cite{ivanov1,ivanov2,ivanov3,ivanov4,ivanov5,ivanov7,ivanov8,ivanov9,ivanov10,ivanov11}. Yet, the $N = 2$, $d=1$ SUSY quantum mechanics supercharges
do not generate spacetime supersymmetry. Following the same lines of research, the supersymmetry in
supersymmetric quantum mechanics, does not directly relate fermions and bosons, and does not of course classify these as representations of the Poincare algebra in four dimensions. Furthermore, the SUSY QM supercharges do not generate transformations between fermions and bosons. Nevertheless, these supercharges classify the Hilbert space of quantum states according to the group $Z_2$. In addition,
the supercharges generate transformations between the Witten parity eigenstates, which are two orthogonal eigenstates of the quantum Hamiltonian. In conclusion, the supersymmetric quantum mechanics algebra is not a spacetime supersymmetry, but a quantum algebra that some of the representations of the super-Poincare algebra obey, at least in the theoretical framework we are working in this paper.

\subsection{$N=2$, $d=1$ SUSY QM Algebra of the Bosonic Fluctuations}

The bosonic zero modes equations for $\kappa =0$ and $A^0=N=0$ are cast in the following form:
\begin{align}\label{selfdualfluctuat1fd}
&(D_1+iD_2)\delta \phi-ie\phi (\delta A_1+i \delta A_2)=0
\\ \notag &(\partial_1-\partial_2)(\delta A_1+ i\delta A_2)+2ie\phi^*\delta \phi =0
\end{align}
Note that the above equations become identical to equations (\ref{selfdualeqns}), if we substitute:
\begin{equation}\label{biossubs}
\psi_{\downarrow}=\delta \phi,{\,}{\,}{\,}\chi_{\uparrow}=\frac{i}{\sqrt{2}}(\delta A_1+i\delta A_2)
\end{equation}
Hence, we conclude that the number of the zero modes corresponding to equation (\ref{selfdualfluctuat1fd}), is equal to the total number of zero modes corresponding to equation (\ref{selfdualeqns}), that is $2n$. Clearly, we can understand that, as in the fermionic case, a $N=2$ SUSY quantum mechanical algebra underlies the bosonic system as well. The structure of the algebra must clearly be the same as the fermionic one, with the difference that the corresponding Hilbert space vectors should be different. Indeed, equations (\ref{selfdualfluctuat1fd}) can be written in the following form:
\begin{equation}\label{transfrey}
\mathcal{D}_{LG}'|\Phi_{LG}\rangle=0
\end{equation}
where, the operator $\mathcal{D}_{LG}'$ is defined to be:
\begin{equation}\label{susyqmrrtyurn567m}
\mathcal{D}_{LG}'=\left(%
\begin{array}{cc}
 D_1+iD_2 & -\sqrt{2}e\phi
 \\ -\sqrt{2}\phi^* & \partial_1-i\partial_2\\
\end{array}%
\right)
\end{equation}
and is considered to act on the vector:
\begin{equation}\label{ait3urtu4e1}
|\Phi_{LG}\rangle =\left(%
\begin{array}{c}
  \delta \phi \\
  \frac{i}{\sqrt{2}}(\delta A_1+i\delta A_2) \\
\end{array}%
\right).
\end{equation}
Therefore, we arrive to the conclusion that:
\begin{equation}\label{dimektriier}
\mathrm{dim}{\,}\mathrm{ker}\mathcal{D}_{LG}'=2n
\end{equation}
Furthermore, as in the fermionic case, we have for the operator  ${\mathcal{D}_{LG}'}^{\dag}$,
\begin{equation}\label{dimekegkjl1r11}
\mathrm{dim}{\,}\mathrm{ker}{\mathcal{D}_{LG}'}^{\dag}=0
\end{equation}
The operators ${\mathcal{D}_{LG}'}$ and ${\mathcal{D}_{LG}'}^{\dag}$ are Fredholm, and therefore any which operator constructed from these, is also Fredholm. The supercharges and the Hamiltonian, that constitute the $N=2$, $d=1$ algebra in the bosonic case, are:
\begin{equation}\label{sgggggg7}
\mathcal{Q}_{LG}'=\bigg{(}\begin{array}{ccc}
  0 & \mathcal{D}_{LG}' \\
  0 & 0  \\
\end{array}\bigg{)},{\,}{\,}{\,}{{\mathcal{Q}'}^{\dag}}_{LG}=\bigg{(}\begin{array}{ccc}
  0 & 0 \\
  {\mathcal{D}'}_{LG}^{\dag} & 0  \\
\end{array}\bigg{)},{\,}{\,}{\,}\mathcal{H}_{LG}'=\bigg{(}\begin{array}{ccc}
 \mathcal{D}_{LG}\mathcal{D}_{LG}^{\dag} & 0 \\
  0 & \mathcal{D}_{LG}^{\dag}\mathcal{D}_{LG}  \\
\end{array}\bigg{)}
\end{equation}

\noindent These three elements of the algebra, satisfy the $d=1$ SUSY algebra:
\begin{equation}\label{relationsforsusy}
\{\mathcal{Q}_{LG}',{\mathcal{Q}'}^{\dag}_{LG}\}=\mathcal{H}_{LG}'{\,}{\,},\mathcal{Q}_{LG}^2=0,{\,}{\,}{\mathcal{Q}_{LG}^{\dag}}^2=0
\end{equation}
Supersymmetry is unbroken, since the corresponding Witten index $\Delta '$ is a non-zero integer, in this case too. Indeed:
\begin{equation}\label{wittindexfgo}
\Delta '= -2n
\end{equation}
Interestingly enough, we found that the bosonic fluctuations in the self dual Landau-Ginzburg vortex background, are related to a $N=2$ SUSY quantum mechanics algebra, which is identical to the fermionic SUSY quantum mechanics algebra, that we came across earlier. It stands to reason to ask if there is a more rich symmetry structure underlying the fermion-boson system, apart from these two $N=2$ algebras. The answer lies in the affirmative. In order to see this, we compute the following commutation and anti-commutation relations:
\begin{align}\label{commutatorsanticomm}
&\{{{\mathcal{Q}'}_{LG}},{{\mathcal{Q}'}_{LG}}^{\dag}\}=2\mathcal{H},{\,}\{{{\mathcal{Q}}_{LG}},{{\mathcal{Q}}_{LG}}^{\dag}\}=2\mathcal{H},{\,}\{{{\mathcal{Q}}_{LG}},{{\mathcal{Q}}_{LG}}\}=0,{\,}\{{{\mathcal{Q}'}_{LG}},{{\mathcal{Q}'}_{LG}}\}=0,{\,}{\,}\\
\notag & \{{{\mathcal{Q}}_{LG}},{{\mathcal{Q}'}_{LG}}^{\dag}\}=\mathcal{Z},{\,}\{{{\mathcal{Q}'}_{LG}},{{\mathcal{Q}}_{LG}}^{\dag}\}=\mathcal{Z},{\,}\\ \notag
&\{{{\mathcal{Q}'}_{LG}}^{\dag},{{\mathcal{Q}'}_{LG}}^{\dag}\}=0,\{{{\mathcal{Q}}_{LG}}^{\dag},{{\mathcal{Q}}_{LG}}^{\dag}\}=0,{\,}\{{{\mathcal{Q}}_{LG}}^{\dag},{{\mathcal{Q}'}_{LG}}^{\dag}\}=0,{\,}\{{{\mathcal{Q}}_{LG}},{{\mathcal{Q}'}_{LG}}\}=0{\,}\\
\notag
&[{{\mathcal{Q}'}_{LG}},{{\mathcal{Q}}_{LG}}]=0,[{{\mathcal{Q}}_{LG}}^{\dag},{{\mathcal{Q}'}_{LG}}^{\dag}]=0,{\,}[{{\mathcal{Q}'}_{LG}},{{\mathcal{Q}'}_{LG}}]=0{\,}[{{\mathcal{Q}'}_{LG}}^{\dag},{{\mathcal{Q}'}_{LG}}^{\dag}]=0,{\,}\\
\notag &
[{\mathcal{H}'}_{LG},{{\mathcal{Q}'}_{LG}}]=0,{\,}[{\mathcal{H}'}_{LG},{{\mathcal{Q}'}_{LG}}^{\dag}]=0,{\,}[\mathcal{H}_{LG},{{\mathcal{Q}}_{LG}}^{\dag}]=0,{\,}[\mathcal{H}_{LG},{{\mathcal{Q}}_{LG}}]=0,{\,}
\end{align}
with $\mathcal{Z}$:
\begin{equation}\label{zcentralcharge}
\mathcal{Z}=2\mathcal{H}_{LG}=2{\mathcal{H}'}_{LG}
\end{equation}
The element $\mathcal{Z}$ commutes with all the elements of the two algebras, namely the
supercharges ${{\mathcal{Q}}_{LG}},{{\mathcal{Q}'}_{LG}}$, their conjugates ${{\mathcal{Q}}_{LG}}^{\dag},{{\mathcal{Q}'}_{LG}}^{\dag}$ and
finally the Hamiltonians, $\mathcal{H}=\mathcal{H}_{LG}={\mathcal{H}'}_{LG}$.
The relations
(\ref{commutatorsanticomm}) describe an $N=4$ supersymmetric quantum mechanics
algebra with central charge $\mathcal{Z}$, which is the Hamiltonian of each $N=2$ subsystem. Indeed, for an $N=4$ algebra we have:
\begin{align}\label{n4algbe}
&\{Q_i,Q_j^{\dag}\}=2\delta_i^jH+Z_{ij},{\,}{\,}i=1,2 \\ \notag &
\{Q_i,Q_j\}=0,{\,}{\,}\{Q_i^{\dag},Q_j^{\dag}\}=0
\end{align}
Clearly the algebra (\ref{commutatorsanticomm}) possesses two central charges which
are equal. Particularly the central charges are $Z_{12}=Z_{21}=\mathcal{Z}$.

\noindent Obviously, such a structure is particularly interesting, since the fermions and the fluctuations of the bosonic fields, that are connected with a $N=2$ spacetime supersymmetric algebra, in $(2+1)$-dimensions, constitute the Hilbert space of an $N=4$ $d=1$ supersymmetric quantum mechanics algebra.

\noindent The $N=4$ supersymmetric algebra is very
important in string theory, since extended (with $N=4,6...$) supersymmetric quantum
mechanics models are the resulting models from the dimensional reduction to one
(temporal) dimension of $N=2$ and $N=1$ Super-Yang Mills theories, as we already mentioned \cite{ivanovbook}.  Moreover, extended supersymmetries serve as super-extensions of
integrable models like Calogero-Moser
systems, Landau-type models \cite{ivanovbook}.  But one of the most salient feature of the $N=4$ extended
supersymmetric quantum algebra is that it can be connected to a generalized harmonic
superspace
\cite{ivanov1,ivanov2,ivanov3,ivanov4,ivanov5,ivanov7,ivanov8,ivanov9,ivanov10,ivanov11},
with the latter being a useful tool for $N\geq 4$ supersymmetric model building.
These harmonic space structures are linked to supersymmetric linear sigma models defined
in the target space. In addition, the harmonic variables give rise to target space
harmonics.

\subsubsection{$SO(2)$ Quantum States Symmetry and the Metric of Vortices}

We shall now address an issue related to the metric of the space of solutions corresponding to the vortices. The dynamics of slowly moving vortices may be described by geodesic motion on the manifold of static bosonic soliton solutions, where the metric on the manifold is described by the kinetic energy of the fields. The metric looks like \cite{rebbi,ruback}:
\begin{equation}\label{metrichjfh}
\mathrm{d}s^2=\int \mathrm{d}^2x\Big{(}\langle\mathrm{d}A_i\arrowvert\mathrm{d}A_i\rangle+\langle\mathrm{d}\phi_i\arrowvert\mathrm{d}\phi_i\rangle\Big{)}
\end{equation}
Hence, we can see that the zero mode bosonic fluctuations can sufficiently describe the dynamics of the vortices and also provide a consistent quantum mechanical description of the slowly moving vortices \cite{lee,ruback,rebbi,samols}. If the motion is not slow, then these fluctuations would produce massive bosonic excitations on the $2n$ moduli space manifold of the Higgs solutions \cite{ruback}, but for sufficiently slow motion, these can be disregarded. In addition, the metric has an $SO(2)$ symmetry \cite{samols} which finds a physical explanation to the fact that the system of vortices is unchanged by a constant rotation of all the velocity vectors. It is intriguingly interesting that the fluctuations of the bosonic fields are the Hilbert space states of an $N=2$ supersymmetric quantum mechanics algebra. This is important, since it provides a theoretical framework and helps us to further enlighten the quantum theory of abelian vortices, to which theory, the bosonic fluctuations are
collective coordinates associated with the vortices \cite{lee}. In reference to the $SO(2)$ symmetry, the $N=2$ supersymmetric quantum mechanics algebra, to which the bosonic fluctuations are the Hilbert space states, provides the complex supercharges with a global $U(1)$ symmetry, which passes to the eigenstates \cite{susyqm,susyqm1}. Indeed, the superalgebra and the Hamiltonian are
invariant under the transformations:
\begin{align}\label{transformationu1}
& \mathcal{Q}_{LG_1}=e^{-ia}\mathcal{Q'}_{LG}, {\,}{\,}{\,}{\,}{\,}{\,}{\,}{\,}
{\,}{\,}\mathcal{Q}_{LG_1}^{\dag}=e^{ia}\mathcal{Q'}_{LG}
\end{align}
Thereby, each quantum system is invariant under a global $R$-symmetry of the
form of a global-$U(1)$. The global
$U(1)$ symmetry is also a symmetry of the Hilbert states
corresponding to the spaces $\mathcal{H}^{+}$,
$\mathcal{H}^{-}$. Let, $\psi^{+}_{LG}$ and
$\psi^{-}_{LG}$ denote the Hilbert states corresponding to the
spaces $\mathcal{H}^{+}$ and $\mathcal{H}^{-}$. The
$U(1)$ transformation of the quantum Hilbert space states is equal to,
\begin{equation}
\psi^{'+}_{LG}=e^{-i\beta_{+}}\psi^{+}_{LG},
{\,}{\,}{\,}{\,}{\,}{\,}{\,}{\,}
{\,}{\,}\psi^{'-}_{LG}=e^{-i\beta_{-}}\psi^{-}_{LG}
\end{equation}
The parameters $\beta_{+}$ and $\beta_{-}$ are global
parameters so that their relation to $a$ is $a=\beta_{+}-\beta_{-}$. The connection of this global $U(1)$, to the aforementioned $SO(2)$ symmetry of the metric is obvious, since the two groups are homomorphic, but we can make it more transparent, if we write the complex SUSY quantum mechanics supercharges, in terms of real supercharges. Indeed, we can write:
\begin{equation}\label{superteransfgit}
\mathcal{Q'}_{LG}=\frac{1}{\sqrt{2}}\Big{(}Q_1+iQ_2\Big{)},{\,}{\,}{\,}\mathcal{Q'}_{LG}^{\dag}=\frac{1}{\sqrt{2}}\Big{(}Q_1-iQ_2\Big{)}
\end{equation}
where $Q_1,Q_2$ are real supercharges. Hence the global $U(1)$ transformation of relation (\ref{transformationu1}) becomes:
\begin{equation}\label{extrasrttho2}
\left(%
\begin{array}{c}
 Q_1'
 \\ Q_2'\\
\end{array}%
\right)=\left(%
\begin{array}{cc}
 \cos a & \sin a
 \\ -\sin a & \cos a\\
\end{array}%
\right){\,}\left(%
\begin{array}{c}
 Q_1
 \\ Q_2\\
\end{array}%
\right)
\end{equation}
Obviously, this transformation holds for the fluctuations of the bosonic fields, that is:
\begin{equation}\label{extrahrso2}
\left(%
\begin{array}{c}
 \delta \phi '\\
 \\ \frac{i}{\sqrt{2}}(\delta A_1+i\delta A_2)'\\
\end{array}%
\right) =\left(%
\begin{array}{cc}
 \cos a & \sin a
 \\ -\sin a & \cos a\\
\end{array}%
\right){\,}\left (\begin{array}{c}
 \delta \phi \\
 \\ \frac{i}{\sqrt{2}}(\delta A_1+i\delta A_2)\\
\end{array} \right)
\end{equation}
Therefore, the $SO(2)$ symmetry of the metric of the fluctuations (\ref{metrichjfh}) and the $SO(2)$ symmetry of the supercharges of the corresponding quantum Hilbert states (\ref{extrahrso2}) certainly have the same origin. Hence, this supersymmetric quantum mechanics algebra of the bosonic fluctuations is an interesting ingredient of the quantum theory of the self dual vortices, since it provides information for the theoretical background of the theory.

\subsubsection{$N=2$, $d=1$ SUSY QM and Massive Bosonic Fluctuations}

\noindent The implications of the unbroken $N=2$ supersymmetric quantum mechanics algebra to the non-zero modes eigenfunctions, are quite interesting, since $d=1$ supersymmetry provides a natural grading to the excited quantum states of the quantum system and additionally a relationship between the two graded types of eigenfunctions. Particularly, the non-zero modes quantum states, that is the eigenfunctions of the quantum Hamiltonian, are classified according to their Witten parity to W-positive and W-negative states,
\begin{equation}\label{cllass}
P^{\pm}|\Psi^{\pm}_E\rangle =\pm |\Psi^{\pm}_E\rangle
\end{equation}
where the index $E$ indicates that the above states are not zero modes. Using the representation (\ref{phi5})
\begin{equation}\label{phi5}
|\Psi^{+}_E\rangle =\left(%
\begin{array}{c}
  |\psi^{+}_E\rangle \\
  0 \\
\end{array}%
\right),{\,}{\,}{\,}
|\Psi^{-}_E\rangle =\left(%
\begin{array}{c}
  0 \\
  \psi^{-}_E\rangle \\
\end{array}%
\right)
\end{equation}
where the eigenfunctions $\psi^{\pm}_E$ belong to the graded subspace of the total Hilbert space, that is $\psi^{\pm}_E$ $\in$ $\mathcal{H}^{\pm}$. Note that the action of the supercharges on the graded subspaces $\mathcal{H}^{\pm}$, is as follows:
\begin{equation}\label{fhghjjffj}
\mathcal{Q}_{LG}'\mathcal{H}^{-}\subset \mathcal{H}^{+},{\,}{\,}{\,}{\mathcal{Q}_{LG}'}^{\dag}\mathcal{H}^{+}\subset \mathcal{H}^{-}
\end{equation}
Particularly, in reference to the above relation, the exact relationship between the eigenfunctions is:
\begin{equation}\label{dfgeigen}
{\mathcal{Q}_{LG}'}^{\dag}|\Psi^{+}_E\rangle =\sqrt{E}  |\Psi^{-}_E\rangle, {\,}{\,}{\,}{\mathcal{Q}_{LG}'}|\Psi^{-}_E\rangle =\sqrt{E}  |\Psi^{+}_E\rangle
\end{equation}
with $E$ the non-zero eigenvalue of the Hamiltonian, that is:
\begin{equation}\label{}
\mathcal{H}_{LG}'|\Psi^{\pm}_E\rangle = E |\Psi^{\pm}_E\rangle
\end{equation}
From the above equations naturally follows that the Hamiltonians of the graded subspaces of the total Hilbert space $\mathcal{H}$ are isospectral, that is:
\begin{equation}\label{isosp}
\mathrm{spec}(\mathcal{H}_+)/\{0\}\equiv \mathrm{spec}(\mathcal{H}_
{-})/\{0\}
\end{equation}
Hence, the following relations hold for the operators $\mathcal{D}_{LG}$ and $\mathcal{D}_{LG}^{\dag}$, as a direct consequence of (\ref{dfgeigen}),
\begin{equation}\label{movingframe}
\mathcal{D}_{LG} |\psi^{-}_E\rangle =\sqrt{E} |\psi^{+}_E\rangle ,{\,}{\,}{\,}\mathcal{D}_{LG}^{\dag}|\psi^{+}_E\rangle = \sqrt{E} |\psi^{-}_E\rangle
\end{equation}
Thereby, although the kernel of the operator $\mathcal{D}_{LG}^{\dag}$ is null, the non-zero modes (excited states) spectrum of this operator exist. Moreover, these are equal in number to the non-zero mode spectrum of $\mathcal{D}_{LG}$. We believe that this result is an interesting attribute of the quantum theory of self dual Abelian-Higgs vortices and serves as a small development towards a quantum theory of the self dual vortices.

\section{Soft Supersymmetry Breaking Terms Impact on the Index}

One of the most interesting results that we found in the previous sections, is the fact that the both the bosonic fluctuations and the fermionic fields have the same number of zero modes, a fact that is a consequence of the $N=2$ spacetime supersymmetry of the physical system. It would be interesting to see if the soft breaking of supersymmetry in one of the sectors, has a direct impact on the number of the zero modes and on the underlying supersymmetric quantum mechanics algebras.

\noindent Soft supersymmetry breaking \cite{martin,dimopoulos} is one elegant way to imitate the phenomenological picture of the Standard Model with the superpartners hidden from the low energy spectrum, rendering the resulting theory free of the quadratic divergences of the Standard model. Soft supersymmetry breaking is materialized in Lagrangian terms that have positive mass dimension coupling. We shall make use of the following terms:
\begin{equation}\label{softbreak}
-\mathcal{M}_1\bar{\chi_{\uparrow}}\psi_{\downarrow},{\,}{\,}{\,}-\mathcal{M}_2\bar{\psi_{\downarrow}}\chi_{\uparrow}
\end{equation}
which obviously break supersymmetry when added to the $N=2$ spacetime supersymmetric Lagrangian, because they explicitly give masses to the fermions and not to their supersymmetric superpartners. The addition of these terms modifies the fermionic equations of motion (\ref{fermionsequationsofmotions}), in such a way that the matrix $\mathcal{D}_{LG}$ is modified to be:
\begin{equation}\label{eqndag}
{\mathcal{D}_{s}}=\left(%
\begin{array}{cc}
 D_1+iD_2 & \sqrt{2}e\phi-\mathcal{M}_1
 \\ -\mathcal{M}_2+\sqrt{2}\phi^* & \partial_1-i\partial_2\\
\end{array}%
\right)
\end{equation}
The operator $\mathcal{D}_{s}$ can be written as follows:
\begin{equation}\label{bnewfoirf}
\mathcal{D}_{s}=\mathcal{D}_{LG}+\mathcal{C}
\end{equation}
with $\mathcal{C}$,
\begin{equation}\label{codd}
\mathcal{C}=\left(%
\begin{array}{cc}
 0& -\mathcal{M}_1
 \\ -\mathcal{M}_2 & 0\\
\end{array}%
\right)
\end{equation}
The operator $\mathcal{C}$ is considered to contain non-infinite terms, which is an obvious constraint since the masses originate from Lagrangian mass terms, and consequently it is a bounded operator. Moreover it is an odd matrix. There exists a theorem relating Fredholm operators that are connected the same way as the operators $\mathcal{D}_{s}$ and $\mathcal{D}_{LG}$, which states that:

\begin{itemize}
 \item For $D'=D+C$ with $D$ a trace class operator and $C$ a bounded odd operator, the indices of $D+C$ and $C$ are equal, that is

\begin{equation}\label{indperturbhfgatrn}
\mathrm{ind}_{t}(D+C)=\mathrm{ind}_{t}D
\end{equation}
\end{itemize}
In our case, the condition trace-class is satisfied since the operators $\mathcal{D}_{s}$ and $\mathcal{D}_{LG}$, are Fredholm. In addition, the matrix $\mathcal{C}$ is odd (it commutes with the involution of the $Z_2$ graded supersymmetric algebra), hence the theorem above applies, and we find that:
\begin{equation}\label{indperturbhhgjhjghkjgjfgatrn}
\mathrm{ind}\mathcal{D}_{s}=\mathrm{ind}(\mathcal{D}_{LG}+\mathcal{C})=\mathrm{ind}\mathcal{D}_{LG}
\end{equation}
Hence, we conclude that soft supersymmetry breaking does not affect the Witten index of the underlying supersymmetric algebra, and therefore the $N=2$, $d=1$ SUSY algebra remains unbroken.

\noindent Since the Landau-Ginzburg model vortices solutions are pretty much related to the theoretical framework of the Abelian-Higgs models, and since type II superconductors can be described by Landau-Ginzburg models, it is tempting to think that the soft supersymmetry terms are similar to the impurities terms in anisotropic superconductors \cite{deang}. As we saw, the impurities do not have a direct effect on the zero mode spectrum of the fermionic sector of the $N=2$ Abelian-Higgs models. Moreover, since the number of fermionic zero modes characterize the degeneracy of the solitonic solutions, this number remains invariant, under the soft breaking of spacetime supersymmetry.

\section{The $\kappa \rightarrow \infty$ Case}

The $\kappa \rightarrow \infty$ limit of the model, is qualitatively very much alike to the $\kappa = 0$ case, which we just presented. In the case at hand, the ratio $e^2/\kappa$ remains fixed, and hence the neutral scalar field can be represented in terms of the complex scalar field (the kinetic term of the neutral scalar can be neglected). Moreover, the same arguments apply for the spinor field $\chi$ which can be written in terms of the spinor $\psi$. Therefore, it is possible to write:
\begin{equation}\label{neuscaln2case}
N=-\frac{1}{\kappa}e(\lvert \phi\lvert^2 -v^2),{\,}{\,}{\,}\chi=-\frac{i}{\kappa}\sqrt{2}e\phi^*\psi
\end{equation}
Hence, equation (\ref{eqnmotion}) can take the form:
\begin{equation}\label{neweqnm}
\gamma^iD_i+ie\Big{(}\gamma^0A^0+\frac{e}{\kappa}(3\lvert \phi \lvert^2 -v^2)\Big{)}\psi =0
\end{equation}
Using the conventions (\ref{conventions}), equation (\ref{neweqnm}) can be cast as:
\begin{align}\label{neweqnm}
&(D_1+iD_2)\psi_{\downarrow}+ie\Big{(}A^0+\frac{e}{\kappa}(3\lvert \phi\lvert^2 -v^2)\Big{)}\psi_{\uparrow}=0
\\ \notag & (D_1+iD_2)\psi_{\uparrow}+ie\Big{(}A^0-\frac{e}{\kappa}(3\lvert \phi\lvert^2 -v^2)\Big{)}\psi_{\downarrow}=0
\end{align}
Upon using the equality:
\begin{equation}\label{aoeq}
A^0=\frac{e}{\kappa}(v^2-\lvert \phi \lvert^2)
\end{equation}
which is a direct consequence of (\ref{selfdualeqns}), the equations (\ref{neweqnm}) can be cast in the following form:
\begin{align}\label{neweqnm111}
&(D_1+iD_2)\psi_{\downarrow}+\Big{(}2i\frac{e^2}{\kappa}\phi\Big{)}{\psi'}_{\uparrow}=0
\\ \notag & (\partial_1-i\partial_2){\psi'}_{\uparrow}+2i\frac{e^2}{\kappa}\Big{(}v^2-2\lvert \phi\lvert^2\Big{)}\phi^* \psi_{\downarrow}=0
\end{align}
The zero mode behavior of the above case is similar to the Abelian Higgs model, hence we refrain from going into details. Accordingly, an unbroken $N=2$ supersymmetric quantum mechanics algebra underlies the fermionic sector of the minimal Chern-Simons Higgs Model, with the supercharges and the Hamiltonian in this case being equal to:
\begin{equation}\label{sdss7}
\mathcal{Q}_{CS}=\bigg{(}\begin{array}{ccc}
  0 & \mathcal{D}_{CS} \\
  0 & 0  \\
\end{array}\bigg{)},{\,}{\,}{\,}\mathcal{Q}^{\dag}_{CS}=\bigg{(}\begin{array}{ccc}
  0 & 0 \\
  \mathcal{D}_{CS}^{\dag} & 0  \\
\end{array}\bigg{)},{\,}{\,}{\,}\mathcal{H}_{CS}=\bigg{(}\begin{array}{ccc}
 \mathcal{D}_{CS}\mathcal{D}_{CS}^{\dag} & 0 \\
  0 & \mathcal{D}_{CS}^{\dag}\mathcal{D}_{CS}  \\
\end{array}\bigg{)}
\end{equation}
The operator $\mathcal{D}_{CS}$ is equal to:
\begin{equation}\label{susyqmrngsdg567m}
\mathcal{D}_{CS}=\left(%
\begin{array}{cc}
 D_1+iD_2 & 2i(e^2/\kappa )\phi
 \\ 2i(e^2/\kappa )(v^2-2\lvert \phi\lvert^2 )\phi^* & \partial_1-i\partial_2\\
\end{array}%
\right)
\end{equation}
which acts on the vector:
\begin{equation}\label{aitsgsdg34e1}
|\Psi_{CS}\rangle =\left(%
\begin{array}{c}
  \psi_{\downarrow} \\
  \phi^* \psi_{\uparrow} \\
\end{array}%
\right).
\end{equation}
The Witten index corresponding to the fermionic section is equal to:
\begin{equation}\label{fnwisggsdtten}
\Delta_{CS} =-2n
\end{equation}
and therefore supersymmetry is unbroken. The bosonic sector yields the same set of equations of motion, and thereby the supercharges and the Hamiltonian are the same as the fermionic ones, hence an $N=2$  $d=1$ supersymmetric quantum algebra underlies the bosonic system. In this case however, the quantum Hilbert space vectors, are written in terms of the vectors:
\begin{equation}\label{ait3urtu4e1}
\left(%
\begin{array}{c}
  \delta \phi \\
  -\frac{\kappa}{2e}(\delta A_1+i\delta A_2) \\
\end{array}%
\right).
\end{equation}
Moreover, the fermionic supersymmetry and the bosonic $N=2$, $d=1$ supersymmetry, constitute a $N=4$ $d=1$ supersymmetry with central charge $\mathcal{Z}=2\mathcal{H}_{CS}$. The quantum theory of the self dual vortices has the same attributes, as the Abelian-Higgs model has, which we described in detail in the previous sections. As we already mentioned, the two separate $N=2$, $d=1$ supersymmetries may not necessarily be a result of the fact that the fermionic-bosonic system satisfies an $N=2$ spacetime supersymmetric algebra. Contrastingly, the $N=4$ $d=1$ supersymmetric quantum algebras that the supercharges of the two $N=2$ supersymmetries obey, is an artifact of the $N=2$ spacetime supersymmetry. This will become quantitatively apparent in the next section, where we consider the $N=1$ case.

\subsection{The $N=1$ Global Supersymmetric Model}

As we mentioned earlier, the bosonic part of the $N=1$ and $N=2$ supersymmetric models is the same. Therefore, we can built an $N=2$ supersymmetric quantum mechanics algebra based on the operator $\mathcal{D}_{LG}'$, which for the $\kappa =0$ case is given by relation (\ref{susyqmrrtyurn567m}). However, the fermionic part of the $N=1$ spacetime supersymmetric quantum model is different from the $N=2$ model. Hence the equations of motion for the $N=1$ model and for $\kappa =0$ (for all other values of $\kappa $ similar results hold true), can be directly connected to a $N=2$, $d=1$ algebra. These equations are:
\begin{align}\label{fewtwttwwr1}
&(D_1+iD_2)\psi_{\downarrow}+\sqrt{2}e\phi\chi_{\downarrow}^*=0
\\ \notag & (\partial_1+i\partial_2)\chi_{\downarrow}+\sqrt{2}e\phi\psi_{\downarrow}^*=0
\end{align}
The number of fermionic zero modes is $2n$ in this case too. Hence the Witten index of the underlying $N=2$, $d=1$ algebra corresponding to fermions is non-zero, and thereby, the $N=2$, $d=1$ is unbroken. In the case at hand however, no simple relation exists between the fermions and the bosonic fluctuations. As a result, the system has two completely independent underlying $N=2$, $d=1$ SUSY algebras, which we denote $\mathcal{N}_1$ and $\mathcal{N}_2$. Hence, the system has a total underlying supersymmetric quantum algebra $\mathcal{N}$ of the form:
\begin{equation}\label{undersusy}
\mathcal{N}=\mathcal{N}_1\oplus \mathcal{N}_2
\end{equation}
These two algebras do not combine to form an extended supersymmetric algebra, as in the $N=2$ spacetime case. This supports our conjecture, that the reason behind the centrally extended $N=4$ supersymmetric quantum algebra corresponding to the $N=2$ spacetime case, is this $N=2$ spacetime supersymmetry, which actually directly connects the fermions with the bosonic fluctuations. This fact is materialized in the transformation of the components of the $N=2$ superfields.

\noindent  The $N=2$, $d=1$ supersymmetric quantum algebra does not necessarily originates from a spacetime supersymmetric algebra. Contrastingly, the $N=4$ $d=1$ algebra, actually originates from a spacetime supersymmetric algebra, a fact that is intriguingly interesting. This is because, a quantum structure that is based on a grading of the $N=4$ Hilbert space is directly connected to an $N=2$ super algebra of the three dimensional spacetime. Let us set this in a more formal way. The fermionic solutions of the $N=2$ spacetime algebra, are the sections of the $U(1)$ (Abelian Higgs Model) twisted spin bundle:
\begin{equation}\label{spinbundlef}
E=\mathcal{P}(M)\times S_{spin}
\end{equation}
with $\mathcal{P}(M)$ the double cover of the principal $SO(3)$ bundle of the tangent space $TM$ of the spacetime manifold $M$, and $S_{spin}$ the irreducible representation of the $Spin (3)$. The $N=2$ spacetime algebra directly connects these sections to the bosonic fluctuations of the scalar fields, and also a part of these sections give rise to an underlying $N=4$ supersymmetric quantum mechanics algebra.

\section*{Conclusions}

In this paper we studied supersymmetric quantum mechanical algebras in Chern-Simons abelian gauge theories in $(2+1)$-dimensions with $N=2$ spacetime supersymmetry. We have examined various limits of the models. We analyzed in  detail the $\kappa =0$ case, which makes the theory identical to the $N=2$ Abelian-Higgs model in $(2+1)$-dimensions. In that case, we found that the fermionic system of this model can constitute an unbroken $N=2$, $d=1$ supersymmetric quantum algebra, with zero central charge. Remarkably, the bosonic system, and in particular the bosonic fluctuations, give rise to another unbroken $N=2$, $d=1$ supersymmetric algebra with zero central charge. The $N=2$ spacetime supersymmetry plays a prominent role in this framework, since it directly connects the fermionic fields involved to the $N=2$, $d=1$ superalgebra, to the bosonic fluctuations. As a result, the two superalgebras combine to create an extended $N=4$, $d=1$ superalgebra with non-zero central charge. These qualitative features hold true for the
case $\kappa \neq 0$ and also for the $\kappa \rightarrow \infty$ limit of the theory. We argued that the aforementioned $N=4$ structure is a result that originates from the $N=2$ spacetime supersymmetry of the initial system. We should mention that the $N=2$, $d=1$ supersymmetry is not a spacetime supersymmetry and hence this symmetry is not directly connected to the spacetime $N= 2$ supersymmetry of the system. As we explained in the previous sections, this supersymmetric algebra causes a grading of the Hilbert space of the quantum system in Witten parity even and parity odd states. These are not spacetime fermions or bosons but simply are $Z_2$ graded states. The $N=4$, $d=1$ supersymmetry has its origin in the higher dimensional spacetime supersymmetry, while the $N=2$, $d=1$ supersymmetry has a geometrical origin which we shall not present here, but we will study in another work. The first argument is further supported by the fact that, for the $N=1$ model there also exist two underlying $N=2$, $d=1$ super algebras, but
nonetheless, these do not combine to an $N=4$ superalgebra. 

\noindent We also analyzed what would be the effect of soft supersymmetry breaking in the fermionic sector, on the number of fermionic zero modes. Recall that the fermionic zero modes represent the degeneracy of the soliton states. As a result of the robustness of the index of Fredholm operators against compact perturbations, the soft supersymmetry breaking has no direct effect on the degeneracy of the solitonic states of the system.

\noindent The zero modes of the bosonic fluctuations provide information for the spacetime metric of the manifold, on which the self-dual vortices are free to move slowly. This metric has a rotational $SO(2)$ symmetry. Using the $N=2$, $d=1$ invariance of the supercharges under a global $U(1)$, we came up to the same result for the Hilbert space states of the quantum mechanical system. Moreover, the $N=2$, $d=1$ algebra provides us with information about the non-zero mode spectrum of the bosonic fluctuations. Particularly, these are classified as even and odd Witten parity states and in addition, there exists an isospectrality of the sub-Hamiltonians corresponding to the graded Hilbert subspaces. These results are relevant to the quantum theory of self-dual solitons.

\noindent Finally, we have to mention that $N=2$, $d=1$ superalgebras are usually found in fermionic systems. But as we saw, in the context of $N=2$ spacetime supersymmetry, also bosonic systems possess such a superalgebra. Specifically, the system of bosonic fluctuations constitutes a superalgebra and not the one corresponding to bosonic fields. It would be interesting to find other bosonic physical systems with defects, in which such a superalgebra exists, without the need of a spacetime supersymmetry in the theoretical framework of the theory under study. We hope to address such issues in the future.

\end{document}